\documentclass[a4paper]{article}
\usepackage{graphicx}
\usepackage{amssymb}
\usepackage{latexsym}
\usepackage{graphicx}
\usepackage{amsmath}

\date{}

\title{AdS/CFT duality at strong coupling}

\author{{\bf Matteo Beccaria \, and \,  Carmine Ortix}\\ 
	Dipartimento di Fisica, Universita' di Lecce,\\
        Via Arnesano, 73100 Lecce\\
	INFN, Sezione di Lecce}
\begin{document}
\maketitle

\begin{abstract}
We study the strong coupling limit of AdS/CFT correspondence in the framework of a recently proposed
fermionic formulation of the Bethe Ansatz equations governing the gauge theory anomalous dimensions.
We provide examples of states that do not follow the Gubser-Klebanov-Polyakov law at large 't Hooft coupling $\lambda$, 
in contrast with recent results on the quantum string Bethe equations valid in that regime.
This result indicates that the fermionic construction cannot be trusted at large $\lambda$, although it
remains an efficient tool to compute the weak coupling expansion of anomalous dimensions.
 \end{abstract}

\section{Introduction}

The AdS/CFT correspondence is a duality between string theory on
$AdS_5\times S^5$ and ${\cal N}=4$ super Yang-Mills \cite{Witten:1998qj,Kazakov:2004qf,Klebanov:2000me,Maldacena:1997re,Gubser:1998bc}. If the number of colors
$N\to\infty$ (the planar limit), the correspondence predicts a matching 
between a finite superconformal four dimensional theory and free string theory in a 
peculiar geometry. Both sides of the correspondence are generically not solvable
except in very special limits. In details, the duality map sends special massive string states
to composite operators in the gauge theory. The mass of the string states equals the anomalous dimensions $\Delta$
in the
gauge theory. The non triviality of the two sides of the correspondence implies
the crucial fact that any test of AdS/CFT must explore the non perturbative 
behavior on at least one of the related theories.

In this paper, we discuss the large 't Hooft coupling $\lambda$ limit
where exact calculations can be performed in the string theory that can be 
treated in the supergravity semiclassical approximation. In particular, one should be
able to recover the well known Gubser-Klebanov-Polyakov (GKP) result predicting the scaling law 
 $E\sim 2\,\sqrt{n}\,\lambda^{1/4}$ for the energy of level $n$ massive string states as
$\lambda\to\infty$~\cite{Gubser:1998bc}.

This kind of investigation requires to determine the non perturbative behavior
of anomalous dimensions in the gauge theory. This apparently out of reach task 
has a chance of being performed thanks to the conjectured integrability of  ${\cal N}=4$ SYM \cite{Zarembo:2004,Beisert:2004yq,Beisert:2004ry}. 
It is well known that there are Bethe Ansatz equations computing the correct 
perturbative expansion of the anomalous dimensions of operators with classical dimension $L$
up to the $L$-th loop term \cite{Beisert:2004hm}. 
Technically, the dilatation operator is interpreted as the quantum Hamiltonian acting 
on suitable subsets of gauge invariant operators closed under perturbative renormalization.
Each operator is mapped to the configuration of a spin chain.
In the thermodynamical limit, this allows to derive 
closed exact expressions for the anomalous dimension of particular operators \cite{Zarembo:2005ur,Beisert:2003xu}.
At fixed $L$, it seems hopeless to extend such technique up to the far strong coupling region.

On the other hand, recently new Bethe Ansatz equations have been proposed 
based on a fermionic formulation of the dilatation operator in SYM \cite{Rej:2005qt}. 
The relevant lattice model is a twisted version of the Hubbard model (HM). These equations
have been conjectured to be free of the above {\em wrapping problems}. It is an open
problem to determine whether they capture the strong coupling limit of the gauge theory.

In this paper, we explore these equations and determine in particular the strong 
coupling expansion of two special states, the so-called antiferromagnetic state (AF) \cite{Zarembo:2005ur}
and the folded-string one (FS) \cite{Frolov:2003xy}. These are states placed in the opposite sides of the 
spectrum and are in a sense the simplest to analyze. 

Our results predict an asymptotic scaling $\Delta \sim \sqrt\lambda$ for both states
and apparently are in contrast with the GKP law. This fact opens an interesting  discussion
as we now explain. 

In principle, given a particular gauge invariant composite operator in SYM, it is 
non trivial to find the corresponding massive state in the string theory.
If it is identified, at least in the semiclassical approximation, then it is possible
to compute its energy at strong coupling and compare with the gauge theory result.
In the case of the AF state, the asymptotic law $\Delta\sim\sqrt\lambda$ is conventional 
wisdom. In addition, and as a consequence, it has been proposed a particular semiclassical string 
configuration in the so-called slow string (SS) limit and  conjectured to be dual to the AF state \cite{Roiban}. It has an energy scaling like 
$\sqrt\lambda$, but as we shall point out, with a different prefactor compared with the 
Hubbard model calculation. We believe that this is an indication that the proposed  duality $AF\leftrightarrow SS$ is incorrect. 
Indeed, the recent work \cite{Beccaria:2006td} suggest that the $\sqrt\lambda$ behavior of the AF state is 
totally fake and should not be trusted. This conclusion is based on a 
semi-analytic analysis of the quantum string Bethe Ansatz equations (SBA) valid by construction in the strong coupling limit \cite{Arutyunov:2004}.

In the case of the FS state, the scenario is somewhat different. There is no available solution 
of the SBA equations. On the other hand, the dual string state is identified quite reliably.
As we show, the strong coupling behavior of the semiclassical energy reproduces in this case
the GKP law. Thus, one is led to believe that the unavailable solution to the SBA equations should also 
exhibit a $\lambda^{1/4}$ scaling, again in contrast with the fermion model calculation.

In the following section we present the results obtained in the framework of the HM
leaving to the concluding section a summary of their implications.

\section{The twisted Hubbard model}
As we discussed in the Introduction, the knowledge of the anomalous dimensions in the gauge theory is made possible by the
correspondence between the dilatation operator in the ${\cal N}=4$ SYM and the Hamiltonian of a spin-chain model. In particular, in the SU(2) sector, 
the Hubbard model should predict at all loops and non-perturbatively anomalous dimensions of gauge operators of fixed length $L$ \cite{Rej:2005qt}. 
A key point about the Hubbard model is that at strong 't Hooft coupling $\lambda=g^{2}_{\rm YM} N $, it reduces to a theory of free lattice fermions. 
However one still needs to determine how the weakly coupled gauge theory states flow to lattice fermion states. 
In the following we will consider an half-filled Hubbard Hamiltonian :
\begin{equation*}
H=H_{0}+\frac{\sqrt{\lambda}}{4 \pi}H_{1}
\end{equation*}
where $H_{0}$ contains the on-site Coulomb-type interaction:
$$H_0 = L-\sum_{i=1}^L n_{\uparrow, i}\,n_{\downarrow, i}, \qquad n_{\sigma, i}^{\phantom\dagger} = c^\dagger_{\sigma, i}\,c_{\sigma, i}^{\phantom\dagger}$$
while $H_{1}$ is the free fermion part:
$$H_1 = \sum_{\sigma=\uparrow, \downarrow}\,\left(\sum_{i=1}^{L-1} c^\dagger_{\sigma, i}\,c^{\phantom\dagger}_{\sigma, i+1} + 
e^{i\phi} c^\dagger_{\sigma, L}\,c^{\phantom\dagger}_{\sigma, 1}\right) + \mbox{h.c.}$$ 
The parameter $\phi=\pi/2$ arises from an Aharonov-Bohm flux which is needed to reproduce BDS results \cite{Beisert:2004hm} . In addition, the introduction 
of the twisting phase makes the Hamiltonian invariant under the shift:
\begin{eqnarray}
\label{eq:cyclic}
c_{\sigma, j}&\to& e^{i\phi/L}\ \ \ \ \ \ \,c_{\sigma, j+1},\qquad j=1, \dots, L-1, \nonumber \\ 
c_{\sigma, L}&\to& e^{i\phi(L+1)/L}\,c_{\sigma, 1},
\end{eqnarray}
and related transformation for annihilation operators.
Information about composite operators anomalous dimensions can be derived from the analysis of the full Hubbard model spectrum. We will consider only states 
with total spin zero and invariant under the shift Eq.(\ref{eq:cyclic}). This is a remarkable reduction of the Hilbert space. For example it has been shown 
that in a half-filled $L=8$ lattice the total dimension is reduced from 4900 to 226 \cite{Beccaria:2006aw}. The coupling dependent spectrum flow has been 
evaluated numerically through direct diagonalization method for lattices from $L=4$ \cite{Minahan:2006zi} to $L=8$ \cite{Beccaria:2006aw}. Several crossings 
of the coupling dependent levels have been observed. This apparent violation of the Wigner-Von Neumann non-crossing rule follows directly from nontrivial 
coupling dependent conservation laws and is a characteristic signature of the one-dimensional Hubbard model quantum integrability \cite{Levels}. 

Here a note of caution must be introduced. 
The fermion Hubbard model introduces states with double occupancy that do not have a direct correspondence with the gauge theory composite operators. The role 
of these extra states is still unclear, thus we will consider below the coupling dependent behavior of states that at $\lambda=0$ reduces to states with no 
double occupancy, {\it i.e.} the states that have classical anomalous dimensions $\Delta(\lambda=0)=L$ (the perturbative multiplet in Ref \cite{Beccaria:2006aw}). 
Nevertheless as the 't Hooft coupling increases, these states will mix with all the {\em extra} states with double occupancy.
We will assume an optimistic attitude, waiting for a better understanding of the role of the Hubbard model.  

The direct diagonalization method is clearly turning off for large dimensions of the Hilbert space.
Nevertheless it is possible to investigate longer composite operators using the integrability of the one-dimensional Hubbard model shown in Ref.\cite{LiebWu}. 

After working out the S-matrix one needs to diagonalize the multi-particle system by a nested Bethe Ansatz. The result of this procedure, generalized to the case 
with the twisting phase, yields the following Lieb-Wu equations for an half-filled lattice:
\begin{eqnarray}
L\,q_n &=& 2\pi\,I_n + 2 \sum_{j=1}^{L/2} \tan^{-1}\left[2(u_j-\,\frac{\sqrt{\lambda}}{2 \pi}\,\sin(q_n+\phi))\right], \\
2\pi\,J_k &=&  2 \sum_{j=1}^{L/2} \tan^{-1}(u_k-u_j) -2\sum_{m=1}^L \tan^{-1}\left[2(u_k-\frac{\sqrt{\lambda}}{2 \pi} \,\sin(q_m+\phi))\right], \nonumber
\end{eqnarray}
where $\phi \equiv \pi/(2L)$ and the $I_n$, $J_k$ are the Bethe quantum numbers. The quantum numbers $q_n$ are the Bethe momenta of the particles while the 
$u_i$ are the spin rapidities (Bethe roots) and describe the spin state. 
At $\lambda=0$ the Lieb-Wu equations simplify since the equations for the momenta and the spin rapidities decouple and can be solved successively.
Together with this decoupling, there is a decoupling of the wave function into a charge and a spin part that is determined by the Bethe Ansatz of a 
Heisenberg spin chain. Once the spin rapidities of a particular state are determined in the Heisenberg model, it is possible to follow the evolution 
as the 't Hooft coupling increases, and finally to determine the strong-coupling behavior of any state. 

\section{The AF operator}
\begin{figure}[tbp]
\includegraphics[width=10cm]{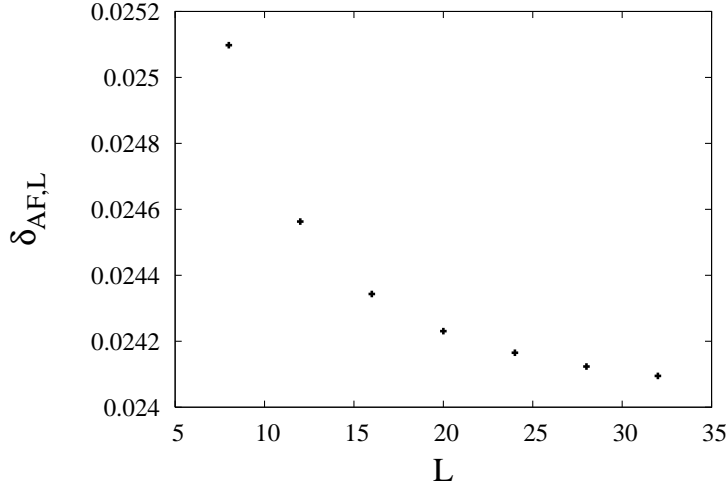}
\caption{Behavior of the next-to subleading term coefficient at strong coupling and finite $L\in 4\mathbb{N}$.}
\label{fig:delta}
\end{figure}

At $\lambda=0$ the state with highest anomalous dimensions corresponds to the ground state of the one-dimensional Heisenberg antiferromagnet. 
The Bethe-Ansatz solution in the thermodynamic limit has been found by Hulth\'en in 1938 \cite{Hulten}. Instead, we determine the Bethe roots for $L$ finite 
and then follow the evolution as the 't Hooft coupling increases by solving step by step the Lieb-Wu equations. 
From the knowledge at every step of the Bethe momenta, we can determine the weak-strong coupling flow of the anomalous dimensions for the 
antiferromagnetic operator. In particular by determining the asymptotic free fermion state at strong coupling, we can provide a series expansion in inverse powers of $\lambda$:
\begin{equation}
\Delta_{\rm AF}^{H}(\lambda,L)=a_{0}(L) \sqrt{\lambda}+a_{1}(L)+a_{2}(L)\dfrac{1}{\sqrt{\lambda}}+\ldots
\label{eq:hubexp}
\end{equation}
\begin{figure}[tbp]
\includegraphics[width=10cm]{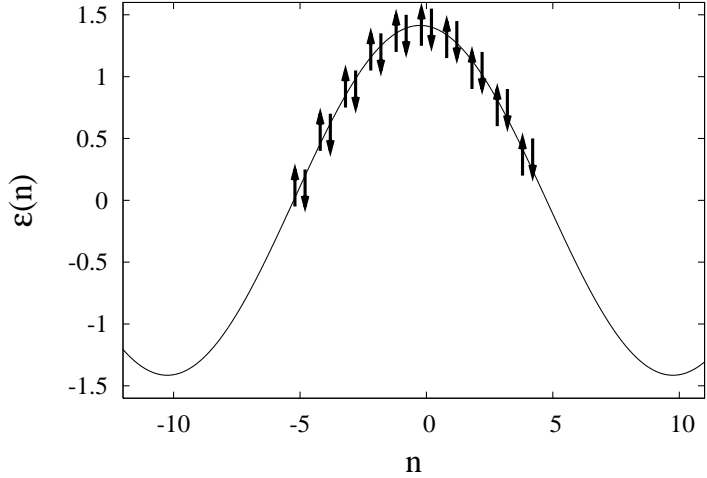}
\caption{Asymptotic free fermion state for the AF operator at $L=20$. }
\label{fig:afmo}
\end{figure}
 We find that the asymptotic free fermion state corresponds to the ground state of the Hubbard hopping term shown in Fig.\ref{fig:afmo}. 
This is the state where all positive energy levels are doubly occupied and thus we have:
$$
|\psi_0^{AF}\rangle = \prod_{n=1}^{L-1} \prod_{\sigma = \uparrow, \downarrow}\,a^\dagger_{\sigma, p_n} |0\rangle
$$
 where $a^\dagger_{\sigma, p_n}$ is the canonical creation operator in the momentum space. 
Since the dispersion relation for the hopping term :
$$\varepsilon_n =2\,\cos\left(\frac{2\pi n}{L} + \frac{\pi}{2L}\right),$$
 we can easily obtain the function $a_{0}(L)$ of the expansion Eq.(\ref{eq:hubexp}): 
$$
a_{0}(L)= \frac{1}{2 \pi} \sin^{-1}\frac{\pi}{2L}
$$
In addition, we can compute analytically $a_{1}(L)$ from first-order perturbation theory. Expressing the Coulombian term $H_{0}$ of the 
Hubbard hamiltonian in momentum space, we obtain:
$$
a_{1}(L) \equiv \langle \psi_0^{AF}|L-\frac{1}{L}\sum_{n ,n^{\prime}, m, m^{\prime}} \delta \left(n-n^{\prime}+m-m^{\prime}\right) 
a^\dagger_{\uparrow, n} a_{\uparrow, n^{\prime}} a^\dagger_{\downarrow, m} a_{\downarrow, m^{\prime}} |\psi_0^{AF}\rangle
$$
It is easy to verify that the latter expression reduces to:
$$
a_{1}(L) \equiv   \langle \psi_0^{AF}|L-\frac{1}{L}\sum_{n , m} a^\dagger_{\uparrow, n} a_{\uparrow, n} a^\dagger_{\downarrow, m} 
a_{\downarrow, m } |\psi_0^{AF}\rangle=L-\frac{1}{L}\left(\frac{L}{2}\right)^{2}=\frac{3}{4}L
$$
Finally the next-to next-to leading order coefficient can be put as:
$$
a_{2}(L)=2 \sqrt{2} \pi \, \delta_{\rm AF,L} \,L
$$
with $\delta_{\rm AF,L}$ numerically evaluated for finite  $L\in 4\mathbb{N}$  up to $L=32$ (see Fig.\ref{fig:delta})
It is interesting to compare our expansion with the prediction of the AF state anomalous dimensions calculated via the BDS
multi-loops Bethe Ansatz equations. In this case a series expansion in inverse power of $L$ is available. Thus:
\begin{equation}
\Delta_{\rm AF}^{BDS}(\lambda, L) =  b_{0}(\lambda) L+ b_{1}(\lambda)+b_{2}(\lambda)\dfrac{1}{L}+ \ldots 
\label{eq:bdsexp}
\end{equation}
The leading order coefficient corresponds to the solution of the BDS Bethe Ansatz in the thermodynamic limit $L \rightarrow \infty$ \cite{Zarembo:2005ur}. It reads:
$$
b_{0}(\lambda)=1+ \frac{\sqrt{\lambda}}{\pi} f\left(\frac{\pi}{\sqrt\lambda}\right)
$$
where 
$$f(x) = \int_0^\infty\frac{dk}{k}\frac{J_0(k)\,J_1(k)}{1+e^{2\,k\,x}},$$
and the $J_{\nu}(x)$ are the Bessel functions of the first kind. 
In order to obtain the next-to and the next-to next-to leading order of the expansion Eq.(\ref{eq:bdsexp}), one has to take into account 
finite size correction to the thermodynamic limit. This can be done by constructing a non-linear integral equation for the 
BDS Bethe Ansatz \cite{Feverati:2006tg}. The subleading order coefficient comes out:
$$
b_{1}(\lambda)=-\dfrac{\lambda}{8 \pi^{2}}\sum_{h=1}^{H}\int_{- \infty}^{\infty} \,dk \,\, 2\pi \,\, e^{i k \, x_{h}}
\dfrac{J_{1}\left(\frac{\sqrt{\lambda}}{2\pi k}\right)}{\sqrt{\lambda} k \cosh{k/2}}
$$
where $H$ and $x_{h}$ represent respectively the number and the position of the Bethe {\em fake} roots. In our case it turns out 
that $H \equiv 0$ since $H=L - 2 M$ and we are considering the case with $M=L/2$ magnons. Thus 
$$b_{1}(\lambda) \equiv 0 $$ 
Contrary the next-to subleading order coefficient results:
$$
b_{2}(\lambda)= \frac{\sqrt{\lambda}}{12} \dfrac{I_{1}\left(\frac{\sqrt{\lambda}}{2}\right)}{I_{0}\left(\frac{\sqrt{\lambda}}{2}\right)},
$$
where $I_{\nu}(x)$ are the modified Bessel functions of the first kind.
We can match the two expressions in the double limit $\lambda, L  \rightarrow \infty $. Within the Hubbard formulation, the anomalous 
dimensions Eq.(\ref{eq:hubexp}) can be expanded at large $L$ to obtain:
\begin{equation}
\Delta_{\rm AF}^{H}= \left[\dfrac{L}{\pi^{2}}+\dfrac{1}{ 24 \,L}+o\left(\dfrac{1}{L}\right)\right] \sqrt{\lambda}+\dfrac{3}{4}L+ 
\left(\delta_{\rm AF, \infty}+ \ldots \right)\dfrac{2 \sqrt{2} \pi }{\sqrt{\lambda}}\,L
\end{equation}
where the coefficient $\delta_{\rm AF, \infty} \sim 0.0240(1)$ has been found from a simple polynomial extrapolation of $\delta_{\rm AF,L}$. 
On the other hand, the BDS prediction can be computed in the strong coupling limit. The subleading order coefficient 
can be expanded asymptotically as \cite{Metzner}:
$$
f(x) = \frac{1}{\pi}-\frac{x}{4}+\sum_{m = 1}^N \mu_m x^{2m} + {\cal O}(x^{2N+2}),
$$
where
$$
\mu_m = \frac{(2m-1)(2^{2m+1}-1)\left[(2m-3)!!\right]^3}{2^{3m-1}(m-1)!}\,\frac{\zeta(2m+1)}{\pi^{2m+1}},\qquad (-1)!!\equiv 1.
$$
and $\zeta$ is the Riemann zeta function. In addition, it is easy to verify that the asymptotic  value of the next-to subleading order coefficient is given simply by:
$$
b_{2}\left(\lambda \rightarrow \infty\right)=\dfrac{\sqrt{\lambda}}{12} 
$$
Inserting the latter expansions in Eq.(\ref{eq:bdsexp}) we find that the two Ansatz have a perfect matching in the terms $\propto L$ but a  
discrepancy of a factor $2$ arises in the term $\sqrt{\lambda}/L$. 
This shows that only in the thermodynamic limit the two different approaches give the same results. The discrepancy in the finite size term
 may be due both to the uncontrolled effect of the {\em wrapping} interactions which start to contribute at the perturbation order $o(\lambda^{L})$ 
in the BDS description and to the different order of limits $\lambda,L \rightarrow \infty$.
More interestingly the leading behavior at strong coupling in both descriptions of the anomalous dimensions scales as $\lambda^{1/2}$. 
 It seems difficult to recover the generic GKP law \cite{Gubser:1998bc}:
$$\Delta \sim 2 \, \sqrt{n} \, \lambda^{1/4}$$ 
It has been thus proposed to identify the dual string state of the antiferromagnetic state with {\em slow-moving} string 
solutions \cite{Roiban} that have the same scaling behavior. However such string states scale as:
$$\Delta \sim \sqrt{\lambda} L$$
This implies that the string solution has a numerical discrepancy $1/\pi^{2}$ in the numerical prefactor with the dual gauge operator. 
This problem has been earlier attributed to the different order of limits since on the string side one first takes the large $\lambda$ 
limit and then consider $L$ large while in the BDS gauge side description one first takes $L$ large and then extrapolate the 
perturbative in $\lambda$ result at strong coupling. We have demonstrated that also in the Hubbard formulation this discrepancy 
survives although one consider the same limit order of the string theory and {\em wrapping} interactions do not contribute.
Along a different route, one can start from the classical string theory at large $\lambda$ and derive string Bethe Ansatz 
equations which are expected to match string calculations including leading quantum corrections. The latter approach has 
been recently studied in Ref.\cite{Beccaria:2006td} and the leading behavior at large $\lambda$ for the antiferromagnetic operator has been shown to reproduce the GKP law.

\section{The FS state}
It is well known that Bethe solution of the Heisenberg antiferromagnet admits the existence of states with spin rapidities that 
acquire an imaginary part. In this case the Bethe wave function is made up of plane waves with complex wave vector. These states 
are usually named `Bethe string solution' and can be identified with quantum Bloch wall state \cite{Dhar,Sutherland}. In the $AdS/CFT$ framework,
 the Bethe string state with all imaginary spin rapidities has the properties \cite{Beisert:2003xu} of the dual to 
semiclassical string  considered in \cite{Frolov:2003qc}. 
We  will instead consider the state with complex spin rapidities that corresponds to the state with lowest anomalous dimensions 
in the perturbative multiplet {\it i.e.} the lightest state. For this state, it has been shown that the Bethe roots (the spin rapidities) 
condense in the thermodynamic limit in two symmetrical curves in the complex plane \cite{Beisert:2003xu} and it is thus named the 
{\em double contour} solution. Its string state dual is nothing but the folded string state described in Ref.\cite{Frolov:2003xy}

\begin{figure}[tbp]
\includegraphics[width=10cm]{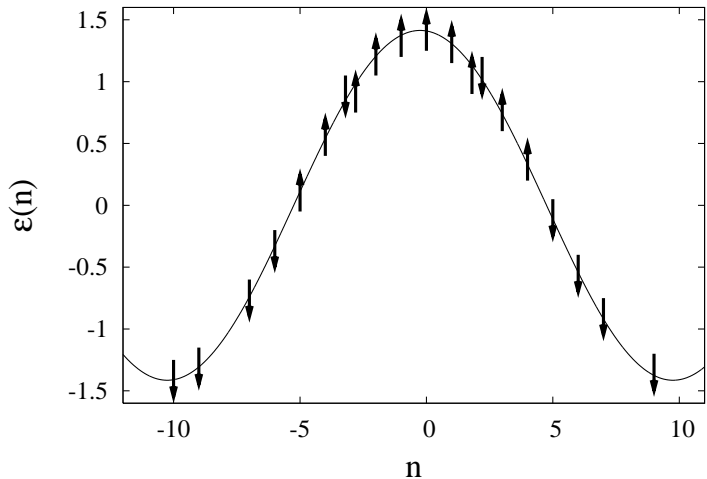}
\caption{Free fermion state for the FS. The modes $n=\pm 8$ are empty. We show only a particular spin configuration for the singly occupied levels.}
\label{fig:fsmo}
\end{figure}

As for the antiferromagnetic operator, we determine at $g=0$ the spin rapidities considering lattices with $L=12,20$ and 
determined step by step the evolution  at strong coupling of $u_i$ as well as of the Bethe momenta $q_n$. 
For both lattices we find that in the limit $g \rightarrow \infty$, the Bethe momenta flow to a state, shown in 
Fig.(\ref{fig:fsmo}), where the positive energy levels with mode numbers $n=k, L-k-1$ are doubly occupied and the 
negative mirror levels are empties. The other positive levels are singly occupied as well as their negative mirror 
levels, thus they do not contribute to the leading term of the state energy. Since the results are completely 
identical for the two lattices we conjecture that for all $L=4(2k+1)$ the pattern will be identical. From the 
dispersion relation for the hopping term of the Hamiltonian the leading term of the anomalous dimensions at strong coupling results:
\begin{equation}
\Delta_{\rm FS}^{0}(\lambda, L) = \dfrac{\sqrt{2}}{\pi} \,\cos\frac{\pi}{2L}\,\sqrt{\lambda}
\end{equation}
The next-to leading term is computed from perturbation theory. It is easy to verify that the non-vanishing matrix elements of $ \langle \psi_0^{FS}|H_{0}|\psi_0^{FS}\rangle$ can be cast as:
\begin{equation}
\Delta_{\rm FS}^{1}(g, L)= \langle \psi_0^{FS}|L-\dfrac{1}{L}\sum_{n , m} a^\dagger_{\uparrow, n} a_{\uparrow, n} 
a^\dagger_{\downarrow, m} a_{\downarrow, m }- \dfrac{1}{L}\sum_{n,m} a^\dagger_{\uparrow, n} a_{\uparrow, m} a^\dagger_{\downarrow, m} a_{\downarrow, n }|\psi_0^{FS}\rangle
\end{equation}
The first two terms have been computed for the AF operator. The third term, contrary is a  spin-flip
 that gives non zero contribution only between singly occupied mirror levels. Thus we have:
\begin{equation}
\Delta_{\rm FS}^{1}(g, L)=L-\dfrac{1}{L}\left[\left(\dfrac{L}{2}\right)^{2}+\left(\dfrac{L}{2}-2\right)\ \right]
\end{equation} 
Finally the series expansion for the folded string anomalous dimensions can be put as:
\begin{equation}
\Delta_{\rm FS}(g, L) = \dfrac{\sqrt{2}}{\pi}\,\cos\frac{\pi}{2L}\,\sqrt{\lambda} + \frac{3L^2-2L+8}{4L} + \delta_{\lambda, L}\,\frac{2 \sqrt{2} \pi}{g} \, L+\cdots,
\end{equation}
Similarly to the antiferromagnetic state, the next-to next-to leading order coefficient $\delta_{\lambda, L}$ can be computed from the numerical exploration of the Lieb-Wu equation. We find:
\begin{eqnarray*}
\delta_{\lambda, 12}&=&0.597(1) \\
\delta_{\lambda, 20}&=&0.953(1)
\end{eqnarray*}

On the other hand, the energy of a folded string with angular momentum $J=L/2$ reads \cite{Plefka:2005bk}:
\begin{equation}
\lim_{L\to \infty}\frac{\Delta_{\rm FS}(\lambda' L^2, L)}{L} \equiv F(\lambda') = \frac{1}{2} K(q)\left[\frac{4q\lambda'}{\pi^2} + \frac{1}{E(q)^2}\right]^{1/2},
\end{equation}
where $q = q(\lambda')\leq 1 $ is the solution of 
\begin{equation}
\frac{4\lambda'}{\pi^2} = \frac{1}{(K(q)-E(q))^2}-\frac{1}{E(q)^2},
\end{equation}
and $K(q)$, $E(q)$ are standard complete elliptic integrals of the first and second kind. 
By expanding at large $\lambda'$, the second equations states: 
$$ q(\lambda')=\frac{2}{\sqrt{\lambda'}}$$
and reinserting in the expression for $\Delta$ we obtain:
$$\lim_{L\to \infty}\frac{\Delta_{\rm FS}(\lambda' L^2, L)}{L}\equiv F(\lambda') \stackrel{\lambda'\rightarrow \infty}{\longrightarrow} \frac{1}{\sqrt{2}}\lambda'^{1/4} $$
Similarly to the AF state we find a natural strong coupling discrepancy between the gauge side and string side. 
The different scaling behavior in $\lambda$ suggests that it seems dangerous to rely on the gauge Bethe Ansatz equations to 
estimate the strong coupling limit of general states.

\section{Conclusions}

To summarize, we have computed in the Hubbard model framework the strong coupling expansion of the 
anomalous dimensions of the AF and FS states. In both cases we find a leading term $\Delta \sim \sqrt\lambda$
at large $\lambda$.
In view of the existing results discussed in the Introduction, we can draw several conclusions.

First, in the case of the AF state, the Hubbard model is totally unable to predict reliably the strong coupling. 
The GKP law is violated whereas there are strong indications in support of it from the exact solution of the 
string Bethe Ansatz equations.

Second, as a consequence, the duality $AF\leftrightarrow SS$ appears to be in quite bad shape.
It is possible that for some gauge theory composite operator the scaling  $\Delta\sim\sqrt\lambda$ holds. However, this does not
seem to be the case, at least  for the AF state. 

Finally, in the case of the FS state, the conclusions are somewhat depending on the (not yet available) solution of the 
string Bethe Ansatz. If GKP scaling were found to hold, then we could conclude that even in this case
the Hubbard model is not predicting the correct strong coupling. Of course, there is also the possibility 
of a violation of the GKP law. However, This option would rule out the duality with the folded string configuration.
In our opinion, this possibility is disfavored due to the perturbative agreement (up to the 3 loop discrepancy)
between the string calculation and the (weak) perturbative expansion of the gauge theory in the BMN limit~\cite{Serban:2004jf}.

\end{document}